\documentclass[aps,prb,groupedaddress,showpacs,nofootinbib]{revtex4-1}
\usepackage{caption}
\usepackage{bm}
\usepackage{amsmath}
\usepackage[cp1251]{inputenc}

\input{epsf}

\begin{document}

%\preprint{Phys.Rev.Lett.}

\title{Size and Shape  Effects in the  Orbital  Magnetization  of TMDs
Monolayers}

\author{ A.V. Chaplik$^{1,2}$ and L.I. Magarill$^{1,2}$}
\affiliation{$^{1}$Institute of Semiconductor Physics, Siberian
Branch of the Russian Academy of Sciences, Novosibirsk, 630090, Russia
\\ $^{2}$ Novosibirsk State University, Novosibirsk, 630090, Russia}

\begin{abstract}
The intrinsic orbital magnetization of a TMD monolayer is usually calculated for a plane unbounded system without mentioning the geometrical shape of samples and boundary conditions (BCs) for electron wave functions. The method of calculation used by many authors (see references below) needs to  account for the contribution of the Berry curvature   also in the case when the system is described by the two-band minimal model \cite{di xiao}. In the present paper, we show that the geometrical and topological properties of the specimen, as well as the BCs, play an important role in the problem of magnetization even for a macroscopic specimen.
\end{abstract}
%%\cite{niu},\cite{di xiao1},\cite{thonhauser},\cite{ceresoli},\cite{di xiao2},\cite{di xiao3},\cite{thonhauser1},\cite{tahir},\cite{di xiao}
\maketitle
\section{Introduction}
 The problem of quantum mechanical calculations of the magnetic moment of a macroscopic object attracted considerable attention in  the  past decade. There are many papers published in the past years \cite{niu, di xiao1, thonhauser, ceresoli, di xiao2, di xiao3, thonhauser1, tahir} where this problem is discussed.  Rich bibliography can be found in the review paper \cite{thonhauser1}.  The overall result of these works reads: for a correct calculation of the orbital magnetization of a large periodic system (like a crystal), it is necessary to take into account the Berry curvature contribution. The latter is determined by $u_{\bf k}({\bf r})$ - the Bloch amplitude which is a part of the total electron wave function $ \psi=  u_{\bf k}({\bf r}) \exp(i{\bf kr}) $   periodically depending on coordinates.
The same approach  was  applied to the calculation of the orbital magnetic moment  in the case of multicomponent  single-particle wave functions  of 2D systems - graphene and TMD  monolayers in the framework of the minimal two-band model  proposed by Xiao et al \cite{di xiao}.  In this model the single electron wave function of a state with a given momentum value ${\bf k}$ is a two-component  spinor  $u_{\bf k}\exp(i{\bf kr})$. The Berry correction to the magnetization is calculated   with the
spinor $u_{\bf k}$ playing the role of Bloch amplitude, though,  in this model, $u_{\bf k}$ does not depend on coordinates (see, e.g., \cite{yang}). All the above-mentioned papers relate to a plane unbounded system, and the geometrical shape of samples and boundary conditions (BCs) for the electron wave functions are not discussed.

    In the present work, we demonstrate that the  consistent application of standard quantum mechanics methods for the minimal model gives correct results coinciding  (where it is possible to compare) with the previously obtained results. However, our way is shorter and simpler and, what is more important, it allows us to account for BC.
     Our proposal is to straightforwardly calculate the diagonal matrix element of the operator  $e[{\bf r, \hat{v}}]/2c$ with proper wave  functions of a finite system of an arbitrary size provided that the wave functions  satisfy certain BCs. However, the result essentially depends on BCs at any sizes of a sample if, naturally, the disorder is ignored and electrons move in the ballistic regime. Besides, the geometrical shape and topological properties of the specimen also essentially affect its intrinsic magnetization, as we will show  below with a few examples.

The Hamiltonian we use is the one of the minimal two-band model \cite{di xiao},\cite{falko},\cite{enaldiev}:
\begin{equation}\label{Hamilt}
\hat{H} = \gamma \mbox{\boldmath{$\sigma$}}_\tau \hat{{\bf p}} + \frac{\Delta}{2} \sigma_z; ~~ \mbox{\boldmath{$\sigma$}}_\tau = (\tau\sigma_x, \sigma_y),
\end{equation}
where $\tau =\pm 1$ is the valley index,   ${\bf \hat{p}}$ is the 2D momentum, $\Delta$ is the energy gap,
and we neglect the spin splitting here for simplicity. The velocity operator $\hat{{\bf v}}$ follows from the Hamiltonian Eq.(\ref{Hamilt}) in accordance with the well-known rules (differentiation of operators by time): $\hat{{\bf v}}= \gamma \mbox{\boldmath{$\sigma$}}_\tau $, where  $\gamma$ is the interband velocity (material parameter). In what follows a few specific examples are considered.

 \section{Rectangle ($0\leq x\leq L_1, 0\leq y\leq L_2$) with zero BC ($\psi_1=0$)}
Two-component eigenfunction ($\psi_1, \psi_2$) is:
\begin{eqnarray} \label{rect_func}
\psi_1= A_{nm}\sin(\frac{\pi nx}{L_1})\sin(\frac{\pi my}{L_2}); \nonumber\\
  \psi_2 = \frac{A_{nm}\gamma}{\Delta/2+E}\bigl(\frac{-i \tau \pi n}{L_1}\cos(\frac{\pi nx}{L_1})\sin(\frac{\pi my}{L_2}) + \frac{\pi m}{L_2}\sin(\frac{\pi nx}{L_1})\cos(\frac{\pi my}{L_2})\bigr),
\end{eqnarray}
where $E$  is the eigenvalue  of the state defined below in Eq.(\ref{A&E}),  $L_1, L_2$ are the rectangle sizes, $A_{nm}$ is the normalization coefficient. For $A_{nm}$ and energy spectrum one can obtain
\begin{eqnarray} \label{A&E}
A_{nm} = \sqrt{\frac{2}{L_1L_2}(1 + \frac{\Delta}{2 |E_{nm}|})}; ~
 E_{nm}^2 = \frac{\Delta^2}{4} + \gamma^2(\frac{\pi^2n^2}{L_1^2} + \frac{\pi^2m^2}{L_2^2}),~ n,m=1,2,.. .
\end{eqnarray}

The functions given by Eq.(\ref{rect_func}) provide obvious BCs: the outward current is zero at any point of the specimen's  boundary (the BC problem is discussed in Section III in more detail).
The partial magnetic moment of the state $(n,m,\tau)$ in the $z$-direction is determined by  operator $e[{\bf r, \hat{v}}]/(2c)$ and equals
\begin{equation}\label{rect_part magn}
M_{nm}^{(\tau)} = \frac{e \gamma^2\tau}{2c E_{nm}}.
\end{equation}
The total magnetic moment of the system for the given valley index \ $\tau$ is $M_{tot}^{(\tau)} = \sum_{nm}M_{nm}^{(\tau)} f(E_{nm})$, where  $f(E_{nm})$ are Fermi occupation numbers. For a sufficiently large specimen,  we replace the sum by the integral over $n$ and $m$ and obtain exactly the same value
 \begin{equation}\label{rect_Mtot}
\frac{M_{tot}^{(\tau)}}{L_1 L_2} = e\tau (E_F - \Delta/2)/(2\pi c)=\frac{\tau e}{2 \pi c}(\sqrt{\frac{\Delta^2}{4}+2\pi  \gamma^2 n_s}-\frac{\Delta}{2}) \nonumber \end{equation} ($n_s $ is the electron concentration)
 that was found in \cite{di xiao3} by more complicate  calculations  using  the Berry phase. This coincidence becomes immediately evident if to calculate the integral with Berry curvature $\Omega$ in Eq.(5) of  \cite{di xiao3}.   If the Fermi energy counted from the conduction band bottom is much less than $\Delta$, a very simple result appears also mentioned in \cite{di xiao3}: one effective magneton per particle (``effective'' because $\gamma^2/\Delta = 1/(2 m^*)$, where $m^*$ is the effective mass). In terms of electron concentration $n_s$, the above used condition reads:  $n_s  \ll \Delta^2/4\pi \gamma^2$. For $MoS_2$ we have $n_s \ll 2\cdot 10^{14}$ cm$^{-2}$ - quite a reasonable limitation for any 2D semiconductor system. The summation over both valleys ($\tau = \pm 1$) gives, of course, zero total magnetization in the equilibrium state.

\section{Disc}
An axially symmetrical  system needs to transform the Hamiltonian (\ref{Hamilt}) into cylindrical coordinates $x=r \cos \varphi, ~y=r \sin \varphi$.
 Then we have:
\begin{eqnarray}
\label{disc_eqs}
(\frac{\Delta}{2} - E)\psi_1 - i \gamma e^{-i\tau \varphi}(\tau\frac{\partial}{\partial r }  - \frac{i}{r}\frac{\partial}{\partial \varphi} )\psi_2 = 0; \nonumber \\
 -i \gamma e^{i\tau \varphi}(\tau\frac{\partial}{\partial r } + \frac{i}{r}\frac{\partial}{\partial \varphi} )\psi_1 - (\frac{\Delta}{2} + E)\psi_2 = 0.
\end{eqnarray}
The system of  Eq.(\ref{disc_eqs}) has the following solutions:
\begin{eqnarray}
\label{disc_sol}
\psi_1 = A \frac{e^{i m\varphi}}{\sqrt{2 \pi}}~ J_m(\kappa r); ~ \psi_2 =  A \frac{e^{i(m+\tau)\varphi}}{\sqrt{2 \pi}} \frac{i \gamma \kappa}{E+\Delta/2}~ J_{m+\tau}(\kappa r),  \end{eqnarray}
where $\kappa^2=(E^2-\Delta^2/4)/\gamma^2, ~~~m=0, \pm 1, \pm 2...,~~ J_m $ is the Bessel function. The solutions of Eq.(\ref{disc_sol}) are regular at $r=0$, while, at $r=R$ (the disk radius), they should obey certain BCs. This issue is discussed below.
The normalization coefficient $A$ is given by
\begin{equation}\label{disc_norm}
  \frac{1}{A_m^2} = \frac{1}{\kappa^2}\int_0^{\kappa R}\Bigl[J_m^2(z) + \bigl(\frac{\gamma\kappa}{E+\Delta/2}\bigr)^2 J_{m+\tau}^2(z)\Bigr]z dz.
    \end{equation}

    The partial magnetic moment is
\begin{eqnarray}\label{disc_partmagn}
   \hat{M}_z=\frac{e\gamma r}{2c} \begin{vmatrix}0 & -i e^{-i\tau \varphi}\\
       i e^{i\tau \varphi} & 0 \end{vmatrix},
    \end{eqnarray}                                                                                                                                             and its average value with  spinor $(\psi_1,~\psi_2)$ has the form:
    \begin{equation}\label{disc_magn}
 (M_z)_{m}^\tau = \frac{e\gamma^2 }{c~\kappa^2}\frac{A_m^2}{E+\Delta/2}\int_0^{\kappa R}J_m(z) J_{m+\tau}(z)z^2 dz. \end{equation}

  By means of one of recurrence relations for the Bessel functions  $ zJ_{m+\tau}(z) - mJ_m(z) = -\tau z J_m'(z)$ we come to the final formula:
     \begin{eqnarray}\label{disc_magn_fin}
 (M_z)_{n,m}^{\tau} = \frac{e\gamma^2 }{c~(E_{n,m}+\Delta/2)} \frac{\int_0^{\kappa_{n,m} R}(m+ \tau)J_m(z)^2z dz-\tau (\kappa_{n,m} R)^2J_m^2(\kappa_{n,m} R)/2}{\int_0^{\kappa_{n,m} R}\Bigl[J_{m}^2(z) + \bigl(\gamma\kappa_{n,m}/(E_{n,m}+\Delta/2)\bigr)^2 J_{m+\tau}^2(z)\Bigr] z dz}.  \end{eqnarray}
 Here $n$ is the radial quantum number, $E_{n,m}$ are energy levels which are determined by BCs,  ~$\kappa_{n,m}=\sqrt{E_{n,m}^2-\Delta^2/4}/\gamma$.

 To find an explicit expression for $(M_z)_{n,m}^\tau$, one has to choose certain BCs. This problem, for a more than one-component wave function, was discussed in a  great number of papers  (see review \cite{volk} and  references therein). Due to the linear character of the wave equations the most general form of BCs has to be a linear combination of $\psi_1$ and $\psi_2$ at the boundaries. Berry and Mondragon showed in \cite{berry} that, for a physically most appropriate situation, the sample is surrounded   by a medium with a very wide forbidden gap (some dielectric or vacuum) so that the electrons, neither of conduction band  nor of the valence band, can escape  the sample; this combination for the disk of radius $R$ reads: $F_1(R)=-i \tau F_2(R)$, where $F_{1,2}$ are the radial parts of  solutions $\psi_{1,2}$. Hence, in our case, one has to solve the equation
\begin{equation}\label{disc_disp}
 J_m(\kappa R) = \tau\frac{\gamma\kappa}{E+\Delta/2}J_{m+\tau}(\kappa R)
\end{equation}
to find the eigenvalue $E_{n,m}$. In the limit case  $n_s \ll \Delta^2/4\pi \gamma^2 \equiv n_0, $ the BCs of Eq.(\ref{disc_disp}) becomes simply $J_m(\kappa R)=0$, and the second terms, both in the numerator and denominator of Eq.(\ref{disc_magn_fin}), can be neglected. Then
\begin{equation}\label{disc_magn_tot}
  (M_z)_{tot}^\tau = \sum_{n,m} f(E_{n,m})(m+\tau)\frac{e\gamma^2}{c\Delta} = \tau\mu_B^* N, \end{equation}
  where $\mu_B^*$ is the effective Bohr magneton,  $N$ is the total number of electrons.
  It seems physically obvious that, for the areal electron concentration, much smaller than $n_0$, the Berry-Mondragon BC not only for a disc, but also for rectangle, as well as for a sample of arbitrary shape, becomes equivalent to the much simpler zero BC.

  Consider now a more general case of not small concentrations when the Fermi energy counted from the c-band bottom is not negligible in comparison with $\Delta$ and the Eq.(\ref{disc_disp}) should be solved without any simplifications. Let the number of electrons in the specimen be large  $(N \gg 1)$. Then the majority of the particles occupy states with large quantum numbers; in other words, $\kappa R \gg 1$ because $\kappa R$ is just the number of the De Broglie half-waves of the radial wave function in the interval  $ [0-R]$. That means we may use  an asymptotic  expression for the Bessel functions. All integrals in the Eq.(\ref{disc_magn_fin}) are evaluated analytically and, after some tedious but simple calculations, we get:
       \begin{equation}\label{disc_magn_asymp}
(M_z)_{n,m}^{(\tau)}=-\frac{e \gamma^2}{c\Delta}(m+\tau \sin^2(x))\cos(2x), ~x = \kappa R - \frac{\pi m}{2} - \frac{\pi}{4}.
    \end{equation}
    After averaging over rapid oscillations stemming, say, from the fluctuations of the  radius $R$ in an array of discs and summation over $m$, we find for the average total magnetic moment of valley $\tau$:
 $(M_z)_{tot}^{(\tau)} = \tau\mu_B^*N /4$, i.e. 4  times less (per one electron)  than in the case of small concentrations.

   Thus, the two-band minimal model predicts a different intrinsic magnetization for TMDs monolayer samples of a different geometry and (via electron concentration) of different BCs. This result can be experimentally checked      if to make the valley populations inequal  when the net magnetization appears (for example, by the circularly polarized light absorption).  Note that the dependence of the TMDs magnetization on the sample geometric shape and on the BCs,  even for macroscopic samples, to our knowledge, has not yet been discussed in the literature.

   \section{1D Ring}
Consider a narrow planar ring shaped area confined by two concentric circumferences of radii $R_1>R_2, ~~R=(R_1 +R_2)/2 \gg R_1-R_2$ and let $(R_1-R_2) \rightarrow 0$. In this  1D ring limit the equations of minimal model give the following:
\begin{eqnarray}
\label{ring_eqs}
(\frac{\Delta}{2} - E)\psi_1 - \frac{\gamma}{R} e^{-i\tau \varphi}\frac{\partial \psi_2}{\partial \varphi} \psi_2 = 0; \nonumber \\
 \frac{\gamma}{R} e^{i\tau \varphi} \frac{\partial \psi_1}{\partial \varphi} \psi_1 - (\frac{\Delta}{2} + E)\psi_2 = 0.
\end{eqnarray}
Then:
\begin{eqnarray}
\label{ring_func}
\psi_1 = \frac{a}{\sqrt{2\pi}} e^{i(j-\tau/2)\varphi},~~~ \psi_2=\frac{b}{\sqrt{2\pi}} e^{i(j+\tau/2)\varphi},\nonumber \\ b=a\frac{i\gamma}{R}\frac{j-\tau/2}{\Delta/2 + E_j}; ~~~ \label{ring_Em}
 E_j^2 = \frac{\Delta^2}{4} + \frac{\gamma^2(j^2-1/4)}{R^2}, \nonumber
\end{eqnarray}
where  ~$ j=\pm 1/2,\pm 3/2 ...$.
 The magnetic moment matrix for the 1D ring is given by Eq.(\ref{disc_partmagn}) with $r=R $ and, for the partial magnetic moment of the $j-$th state, we get:
\begin{equation}\label{ring_magn}
  (M_z)_j^{(\tau)} = \frac{e \gamma^2}{c}\frac{(j-\tau/2)(E_j+\Delta/2)}{(E_j+\Delta/2)^2+(j-\tau/2)^2 \gamma^2/R^2}.
\end{equation}

Using evenness $E_j$ in $j$ and combining the terms with $j$ and $-j$ we find, for the total 1D ring  magnetization, the following:
\begin{equation}\label{ring_magntot}
  (M_z)_{tot}^{(\tau)} = -\sum_{j>0}\frac{\tau e \gamma^2\Delta f(E_j)}{c (\Delta^2+4 \gamma^2 j^2/R^2)} = -\frac{\tau e \gamma R}{2c}\sum_{j=1/2}^{j_{max}} \frac{\delta f(E_j)}{\delta^2+j^2},
\end{equation}
where $\delta =\Delta R/(2 \gamma)$, $j_{max}= (N^{(\tau)}-1)/2$, ~ $N^{(\tau)}$ is the number of electrons per one spin projection in  valley  $\tau$. For all known TMDs the parameter $\delta$ is rather large for realistic values of $R$ (in $MoS_2$, for $R=40$ nm, it is around $10^2$; the numerical data for $\Delta$ and $\gamma$ are taken from \cite{di xiao}). Then  all  terms in the sum change slowly and we can replace the summation by integration. This gives
\begin{equation}\label{Mtot_ring}
 (M_z)_{tot}^{(\tau)}=-\tau\mu_B^* \delta\arctan(\frac{j_{max}}{\delta}) = -\frac{\tau e\gamma  R}{2c}\arctan(\frac{2 \pi \gamma  n_l}{\Delta}).
\end{equation}
Here $n_l$ is the linear electron concentration.

There are two limiting cases:
a) small concentrations, $N^{(\tau)} \ll \delta$,    $(M_z)_{tot}^{(\tau)} = -\tau \mu_B^* N^{(\tau)}/2$ (one half of magneton per particle;
b) very high concentrations, $N^{(\tau)} \gg \delta$, $(M_z)_{tot}^{(\tau)} = -\tau (\mu_B^*/2) \pi \delta$, i.e. – saturation, -{\it the total magnetic moment tends to a large constant with increase in $N^{(\tau)}$}. This unexpected result as well as $ \mu_B^*/2$ instead of one $ \mu_B^*$  per particle, is a unique peculiarity of the TMDs 1D ring.

      For typical TMDs the characteristic magnitude of the linear electron concentration $n_l$, from which the saturation regime begins, is very high: $n_l \sim 10^7$ cm$^{-1}$ for $MoS_2$. However, in gapped graphene, the energy gap depends on the type of substrate and can be made much smaller. If, for example, $\Delta = 0.1$ eV, and $\gamma \simeq 3 \cdot 10^8$ cm/s,   we get for the concentration $n_l = \Delta/(2 \gamma\pi)$ - a quite reasonable value $\simeq 10^5$ cm$^{-1}$. For the ring radius $R=100$ nm the saturation regime starts with around $ 6$ electrons per ring  and the limiting value of the magnetic moment per ring  is $4\pi \mu_B^*$. We guess this makes it quite possible to  observe such an interesting effect in experiments with arrays of narrow graphene-on-substrate rings.

   \section{Torus}
Our last example is a torus, a 2D object for which the problem of BC does not exist at all  (the same as for a 1D ring), and only the  double periodicity of the wave functions  in  angles $\varphi$ and $\psi$ must be ensured (see Fig.1).  Now we use toroidal coordinates and describe the electron position on the torus by means of two sets of orthogonal circumferences: a small circle of radius $r$ and a large one embracing the torus center. These lines are the torus  sections by spheres ($\psi = const$) and half-planes ($\varphi = const$) - coordinate surfaces of  toroidal coordinates. If $R$ is the radius of the central circle (dotted line in Fig.2), then, for a given angle $\psi$, the   radius of the corresponding large circle equals $\rho(\psi) = R + r \cos{\psi}$. Thus, we have the following relations between cartesian coordinates $(x,~ y)$ of a plane system and angles $\varphi$
 and $\psi$: $x \rightarrow (R+r\cos{\psi})\varphi,~~~~~~ y \rightarrow r\psi$.
The length element on the torus is  $dl^2 = (R+r\cos{\psi})^2 d\varphi^2 +r^2 d\psi^2$.
From this equation we find the Lame coefficients, gradient operator components and come to the wave equations:
\begin{eqnarray}
% \nonumber to remove numbering (before each equation)
  (\frac{\Delta}{2} -E)\Psi_1 -i\gamma\Bigl(\frac{\tau}{R+r\cos{\psi}}\frac{\partial}{\partial \varphi} -\frac{i}{r}\frac{\partial}{\partial \psi}\Bigr)\Psi_2 = 0; \nonumber \\
 -i\gamma\Bigl(\frac{\tau}{R+r\cos{\psi}}\frac{\partial}{\partial \varphi} +\frac{i}{r}\frac{\partial}{\partial \psi}\Bigr)\Psi_1 -(\frac{\Delta}{2} +E)\Psi_1= 0.
\end{eqnarray}
Let $\Psi_1= A e^{im\varphi}F(\psi)\sqrt{2 \pi},$  then
\begin{equation}\label{tor_Psi2} \Psi_2 = A \frac{e^{im\varphi}}{\sqrt{2 \pi}}\bigl(\frac{m\tau\gamma}{\rho(\psi)} F(\psi)+ \frac{\gamma}{r} \frac{\partial F(\psi)}{\partial \psi}\bigr)/(\Delta/2+E),
\end{equation}
 where $F(\psi)$ is any periodic  in $\psi$ with the period $2\pi$  solution of the Hill-type equation:
\begin{equation}\label{hill}
\frac{d^2 F(\psi)}{d\psi^2} - \frac{r^2}{\rho^2(\psi)}\bigl(m^2 -m\tau \sin{\psi}\bigr)F(\psi) + \frac{r^2}{\gamma^2}\bigl(E^2-\frac{\Delta^2}{4}\bigr)F(\psi) = 0.
\end{equation}

\begin{figure}[ht]  \label{Fig.1}
\centerline{\epsfysize=5.cm\epsfbox{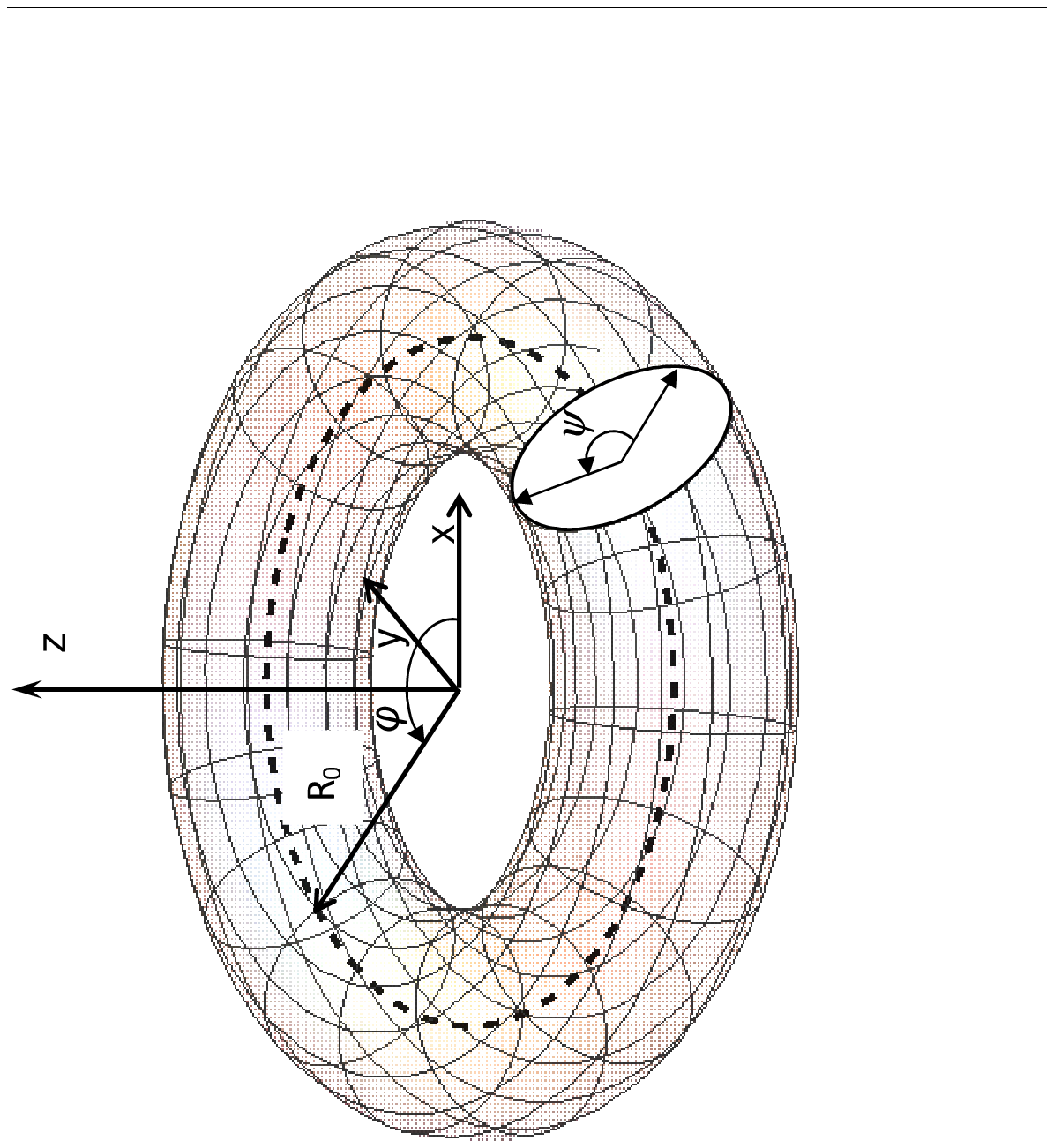}}
\caption{2D system on a torus.}
\end{figure}
The $z$-component  of the magnetic moment operator is  $e \rho^2(\psi) \dot \varphi/2c$,  where $\dot \varphi = \partial \varphi/\partial t$ should be found as $\dot \varphi =i(\hat{H}\varphi - \varphi  \hat{H})$.  This results in
\begin{equation}\label{tor_Mz}
  \hat{M}_z= \frac{e \gamma\tau}{2c}(R + r \cos{\psi})\hat{\sigma}_x.
\end{equation}
Periodic solutions of  the Hill equation Eq.(\ref{hill}) exist, as is known, only  for quantized coefficient values  at $F(\psi)$ in the last term  of Eq.(\ref{hill}) and, hence, for quantized energy values $E_{n,m}$.

The average magnetic moment of the state $(n,~m)$ has the form:
\begin{eqnarray}\label{tor_avM}
  (M_z)_{n,m}^\tau = \frac{e\gamma^2}{c}(\frac{\Delta}{2}+E_{nm})\int_{-\pi}^\pi \Bigl(m F_{n,m}^2(\psi)+\tau \rho(\psi)F_{n,m}(\psi)\frac{\partial F_{n,m}(\psi)}{\partial \psi}\Bigr) \rho(\psi)r d\psi \times \nonumber \\
  \Bigl[\int_{-\pi}^\pi \Bigl(F_{n,m}^2(\psi)\bigl[\bigl(\frac{\Delta}{2}+E_{nm}\bigr)^2 + \bigl(\frac{m\gamma}{\rho(\psi)}\bigr)^2\bigr]+\frac{\gamma^2}{r^2}\bigl(\frac{ \partial F_{n,m}(\psi)}{\partial \psi}\bigr)^2\Bigr)\rho(\psi) rd\psi \Bigr]^{-1}.
\end{eqnarray}
A periodic solution of the Eq.(\ref{hill}) is possible in the form of Fourier series. To describe the torus magnetization  qualitatively, we consider the limiting case of a ``thin'' torus  $R \gg r $ and we put $\rho(\psi)=R_0$. Eq.(\ref{hill}) becomes, in this case, the Mathieu equation (also still difficult for a general solution) and we, again, use  the small parameter $ r/R$. In the zeroth approximation  we have $ F^{''} + \varepsilon F =0$ with $\varepsilon = (E^2- \Delta^2/4)r^2/\gamma^2 - (mr/R)^2$ and the solution $F=\exp{(in\psi)}/\sqrt{2\pi}, ~~n=0, \pm 1,\pm2; ~~E_{n,m}^2 = \Delta^2/4 + (m\gamma/R)^2 +(n\gamma/r)^2)$. Suppose now that the electron concentration is small and the Fermi level lies in the c-band  between $E_{\pm 2,m}$ and $E_{\pm 1,m}$, i.e. only three groups of states  $(0,m), ~(\pm 1,m)$ are occupied for any $m$. Now we have to solve the equation
\begin{equation}\label{tor_simple}
  F^{''} + \frac{m\tau r^2}{R_0^2}\sin{\psi}F + \varepsilon F = 0
\end{equation}
in the three-level approximation. The solution is a linear combination of zero order solutions $F=C_0 + C_{+1}\exp{(i\psi)} + C_{-1}\exp{(-i\psi)}$ and the problem is reduced to the diagonalization of  matrix $3 \times 3$. The results are:
\begin{eqnarray}\label{tor_F}
  &&F_{0} = 1+\frac{2P}{1+\sqrt{1+2 P^2}}\sin{\psi}, ~~~ \varepsilon_0=\frac{1}{2}(1-\sqrt{1+2 P^2}) \approx -\frac{P^2}{2}; \nonumber \\
  &&F_{+1} = \cos{\psi},~~~ \varepsilon_{+1}=1; \nonumber \\
  &&F_{-1} =  1+\frac{2P}{1-\sqrt{1+2 P^2}}\sin{\psi}, ~~~ \varepsilon_{-1}=\frac{1}{2}(1+\sqrt{1+2 P^2}) \approx 1+\frac{P^2}{2}.
\end{eqnarray}
Here $P=m\tau r^2/R_0^2$. Functions $F_{0,\pm 1}$ are not yet normalized solutions. Then we find the partial magnetic moments of the states $0,~1$ and ~$-1$  in accordance  with Eq.(\ref{tor_avM}).  In the leading order of parameter $P \ll 1$ one can obtain for $(M_z)_0$:
\begin{equation}\label{M0}
 (M_z)_0= \frac{me\gamma^2}{2c E_0}\approx m\mu_B^*; ~~~ E_0^2 \approx \frac{\Delta^2}{4} + (\frac{m\gamma}{R_0})^2 +\frac{\varepsilon_0\gamma^2}{r^2} \approx \frac{\Delta^2}{4}.
\end{equation}
  We put $E_0  \approx~ \Delta/2$  in Eq.(\ref{M0})  based on the following considerations: i) the characteristic length $\gamma/\Delta$ for all TMDs  is of the order of
few angstroms, ii) the large torus  radius is, at least, two orders of magnitude higher and
iii) the quantum number $m$ is not extremely large.

The expression for $(M_z)_{+1}$ reads
\begin{eqnarray}\label{M1}
 (M_z)_{+1} = \frac{me\gamma^2}{c \Delta}\bigl(1+\frac{4\gamma^2}{\Delta^2r^2}\bigr); ~~  E_{+1}^2 = \frac{\Delta^2}{4} + (\frac{m\gamma}{R_0})^2 +\frac{\gamma^2}{r^2} \approx \frac{\Delta^2}{4} + \frac{\gamma^2}{r^2}.
\end{eqnarray}
We keep the second term in the brackets, though $\gamma/r$ is still less than $\Delta$ in a more or less real situation.
For example, if one rolls a nanotube with diameter $20$ {\AA} into a torus, the correction to $(M_z)_{+1}$ is approximately $10$\%.

 For the state $(-1)$ we obtain $ E_{-1,m} = \Delta^2/4 + (m\gamma/R_0)^2 +(1+P^2/2)\gamma^2/r^2$, and $(M_z)_{-1}$ differs from $(M_z)_{+1}$ by the order of $P^2$. Thus, for all the three states, the  partial magnetic moment does not depend on the valley index $\tau$ and is proportional to $m$. Hence, the total magnetic moment of each valley equals zero.

 We also checked an alternative geometry on the torus: the $x$-direction in a plane TMDs  layer corresponds to the small torus radius, i.e. $x \rightarrow r \psi$ (we mean that the $x$-direction in the plane is parallel to the line connecting valleys $+1$ and $-1$). In this case, the partial moment of the $n,m$-state does not explicitly contain $\tau$ and is again proportional to $ m$. Thus, the total magnetic moment of any valley vanishes.
\section{Summary}
In all the considered examples we did not come across  any difficulties caused by the integration of  operator $[{\bf r,v}]$ over the  sample area. All factors proportional to the  ``infinite'' sizes are canceled if  properly normalized wave functions are used. The standard quantum mechanical approach leads to quite reasonable results without any additional complications. The total magnetization of a specimen in the absence of external magnetic field equals zero in the equilibrium state but the magnetization of the individual valleys in  TMDs can be nonzero for certain cases. We  considered four different examples and showed that, in the torus the magnetization of each valley separately equals zero, while, in the cases of rectangle, disc and 1D ring,  it has the finite value. For the rectangle and disc, this value is of the order of one effective Bohr magneton per electron in the conduction band if the Fermi energy counted from the band bottom is much smaller than the gap, $E_F - \Delta/2 \ll \Delta$. For the 1D ring in the same  case of small  electron concentration we obtain one half effective magneton per particle. However, all results change drastically if the concentration is not small. For the rectangle the total magnetic moment of each valley is proportional to $E_F - \Delta/2 $, i.e., nonlinearly depends on the number of electrons. In a round disc with $N \gg 1$ the magnetization contains a constant and a part rapidly oscillating as a function of the disc radius; the average magnetic moment equals $1/4$ of an effective magneton per electron. Especially remarkable result is obtained for the 1D ring where the problem allows the  exact analytical solution. Here we again have $\mu_B^*/2$ per particle for small linear concentrations and the saturation of the total magnetic moment if the number of electrons tends to infinity (actually ``zero'' magneton per electron).  Note that all these details, to our best knowledge, have not been discussed yet in the preceding contributions on  the  orbital magnetization of  2D TMDs.

\paragraph*{\bf Acknowledgments.}  This work was  supported  by the  RSF,  grant No 17-12-01039.

\end{document}